\documentclass[11pt,a4paper]{article}
\usepackage[hyperref]{naaclhlt2019}
\usepackage{times}
\usepackage{latexsym}

\usepackage{algorithm}
\usepackage{algpseudocode}
\usepackage{graphicx}
\usepackage{pifont}
\usepackage{program}
\usepackage{multirow}
\usepackage{ragged2e}

\usepackage{url}

\aclfinalcopy 
\title{An Efficient Approach for Super and Nested Term Indexing and Retrieval}

\author{
  Md Faisal Mahbub Chowdhury \\
  IBM Research AI  \\
  {\tt mchowdh@us.ibm.com} \\\And
  Robert Farrell \\
  IBM Research AI  \\
  {\tt robfarr@us.ibm.com}\\
  }

\date{}

\begin{document}
\maketitle
\begin{abstract}
This paper describes a new approach, called \emph{Terminological Bucket Indexing (TBI)}, for efficient indexing and retrieval of both nested and super terms using a single method. We propose a hybrid data structure for facilitating faster indexing building. An evaluation of our approach with respect to widely used existing approaches on several publicly available dataset is provided. Compared to Trie based approaches, TBI provides comparable performance on nested term retrieval and far superior performance on super term retrieval. Compared to traditional hash table, TBI needs 80\% less time for indexing.
\end{abstract}

\section{Introduction}
\label{sec_intro}
A super term is a term that contains another term, e.g. ``Google LLC'' is a super term of the two corresponding nested terms (aka sub-terms) ``Google'' and ``LLC''. Super and nested terms are used in many Natural Language Processing (NLP) applications/tasks, e.g. terminology extraction \cite{frantzi:1998}, taxonomy/ontology population \cite{basili:2003, lassoued:2016}, query expansion/re-formulation \cite{polozov:2012}, query completion, etc.

The task of finding super and nested terms can be defined as follows. Given an input term, retrieve all corresponding super or nested terms from a list of terms (i.e. vocabulary), typically extracted from document corpora. To facilitate such retrieval, one typically indexes the aforementioned vocabulary in a data structure. If the indexing algorithm is inefficient, it can result in substantially more time to build an index for a large vocabulary and can be a bottleneck for NLP applications. 

The most commonly used data structure for term indexing is a key-value store, aka hash table \cite{luhn:1953, knuth:1998}, where key is a term and value is a set of super/nested terms. For example, \citet{zhang:2016} used such an implementation\footnote{https://github.com/ziqizhang/jate}. The indexing time complexity is $O(v^{2})$ where \emph{v} is the size of the vocabulary. Assuming the values in the hash table are super terms (same holds for nested terms), the space complexity is $O(v*s)$ where \texttt{s} is the average number of super terms per term and $v \gg s$. Search time complexities for retrieval of super terms and nested terms (from same hash table) are $O(1)$ and $O(m^2)$ correspondingly, where $m$ is total tokens in the input term, $O(m^2)$ is time complexity for all possible n-gram generation from input term and $v \gg m$.\footnote{Note, the meanings of symbols ($m, v, t, s$ and $L$) used inside the Big O notation are consistent across this paper.}


\citet{lu:2012} described an approach for nested term finding where the indexing is done by loading the vocabulary in a Trie \cite{DeLaBriandais:1959} where words are nodes. The indexing complexity is $O(v*t)$, where $t$ is the average number of words for a term in the vocabulary and $v \gg t$.\footnote{See \texttt{https://tinyurl.com/yask8u7x} for step-by-step in details of \citet{lu:2012}.} The search complexity for nested terms is $O(m^2)$. However, search complexity for super terms is $O(v*t)$, which is very slow.  So, realistically, this approach can be only used for nested term retrieval.

The \citet{aho:1975}, a string search algorithm based on Trie, can be used for nested term finding. The algorithm is widely used in intrusion detection and anti-virus systems, and many other applications that require fast matching against a pre-defined set of strings/terms. It generalizes the  \citet{matiyasevich:1973} and \citet{knuth:1977} algorithms for multiple exact sub-string (i.e. nested term) matching. Time complexity of nested term retrieval is $O(m+L)$ where $L$ is total length of all terms in vocabulary. But it is very slow for super term retrieval (like \citet{lu:2012})  as it is based on Trie.\footnote{Other string search algorithms (e.g.  Karp--Rabin algorithm \cite{karp:1987}) have similar or worse search complexity.}

A somewhat related work by \citet{sun:2008} addressed chemical name indexing and search. For complex chemical names, e.g. ``methylethyl'', they proposed an algorithm to identify substrings, e.g. ``methyl'' and ``ethyl'', that have higher frequency. They proposed to index these substrings instead of all possible substrings. These substrings are similar to nested terms in the chemical domain. The authors did not discuss what algorithm or data structure they used to index the substrings nor whether their algorithm could be used to efficiently retrieve super terms. 

In the remaining of the paper, we describe our proposed approach and empirical observations. The contributions of this paper are following -- \underline{\textbf{(i)}} a new indexing algorithm, \emph{Terminological Bucket Indexing (TBI)}, that uses a hybrid data structure and can be used for both fast super term and nested term retrieval, \underline{\textbf{(ii)}} empirical results demonstrating that TBI reduces indexing time by a minimum 81\% when compared with hash table, and \underline{\textbf{(iii)}} complexity analysis showing that the search time for super and nested term retrieval using the hybrid data structure is the same as a hash table and better than Trie based algorithms.

\begin{table}
\begin{center}
  \begin{tabular}{|c|c|c|c|}
    \hline
    Data & Total & Avg. no. & Avg. no.  \\ 
     &  &  & of  \\ 
     & unique  & of words  & characters  \\ 
     &  terms &  per term &  per term \\ \hline \hline
    Brown & 95,241 & 2.23 & 15.21 \\ \hline
    WSJ & 92,690 & 2.49 & 17.11 \\ \hline
    SWBD & 28,429 & 2.21 & 13.23 \\  \hline
    NYT & 16,242 & 2.01 & 13.27 \\  \hline
    Google w2v & 3 millions & 1.96 & 13.74 \\  \hline
  \end{tabular}
  \caption{Corpora used for the experiments.}
\label{tbl-corpora}
\end{center}
\end{table}

\section{Proposed Approach}

{\large \bf Index Building:} Proposed \emph{TBI} uses a hybrid data structure (henceforth, \textit{term store}) composed of two hash tables. 
$Table\_SuperTerms$ stores terms as keys and set of corresponding super-terms as values. Its space complexity is $O(v*s)$. Here, $v \gg s$. $Table\_Bucket$ has 4 levels as shown below; e.g., \textit{k1}=21, \textit{k2}=4, \textit{k3}=24 and \textit{k4}=7 for term ``\texttt{united states of america}''. \footnote{The ordering, selection and combination of the 4 types of bucket keys were resulted from through checking output accuracy and runtime performance among a wide variety of keys. One can verify accuracy by simply comparing the list of super terms (for every term of a dataset) indexed using the traditional hash table indexing with that inside $Table\_SuperTerms$  using the proposed TBI.}

\begin{figure}[H]
\includegraphics[scale=0.43]{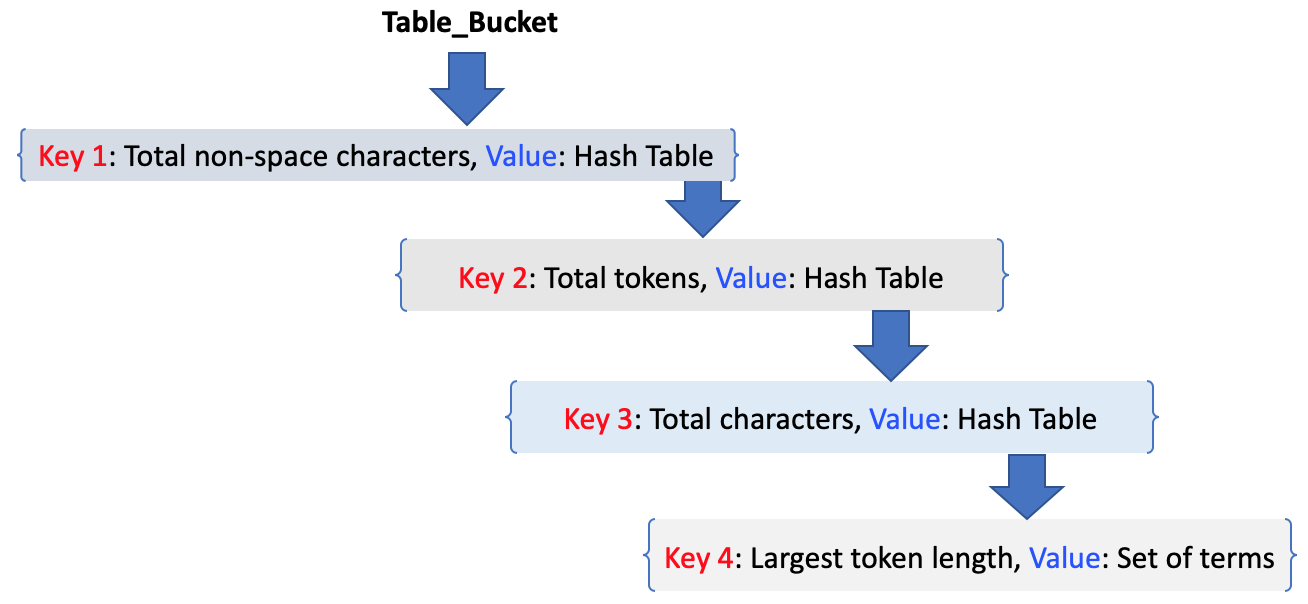}
\end{figure}

The space complexity of $Table\_Bucket$ is $O(v)$ since a term is inserted only once. So, complexity for proposed hybrid \textit{term store} is $O(v*s + v)$. This is a bit higher but still comparable to $O(v*s)$ space complexity of hash table. $Table\_Bucket$ can be discarded after all terms in vocabulary are indexed.

\begin{table*}
\begin{center}
  \begin{tabular}{|c||c|c|c||c|c|c|}
    \hline
     & \multicolumn{3}{|c||}{For nested term} & \multicolumn{3}{|c|}{For super term}  \\ \cline{2-7}
     Data & Aho--Corasick & Lu-Browne & Ours & Aho--Corasick & Lu-Browne & Ours \\ \hline \hline
    Brown & 15.21 & \bf 2.23  & 3.45 & 42433.99  & 33629.19 & \bf 1 \\ \hline
    WSJ & 17.11 & \bf 2.48 & 4.09 & 29170.87 & 22657.57 & \bf 1 \\ \hline
    SWBD & 13.23 & \bf 2.21 & 3.39 & 9556.54 & 12172.2 & \bf 1  \\ \hline
   NYT  & 13.27 & \bf 2.01 & 3.01 & 8673.56 & 6496.11 & \bf 1 \\  \hline
  \end{tabular}
  \caption{Average number of node comparisons needed for term retrieval for different approaches.}
\label{tbl-term-retrieval}
\end{center}
\end{table*}

Algorithm \ref{index_algo} describes proposed TBI. First, the vocabulary is sorted by term length in ascending order (\texttt{line no. 2}). The complexity of this sorting step is  $O(v*log\_v)$ which is very fast. The goal is -- for each (yet to be indexed) \texttt{new\_term} from the vocabulary, efficiently find nested terms from terms that were already indexed (from this sorted vocabulary). Because they are pre-sorted, none of the already indexed terms can be super term of \texttt{new\_term} (as they are shorter or equal). To populate the nested terms for \texttt{new\_term}, the algorithm first looks for the buckets that have terms with lower \textit{`number of non-space characters`} (\texttt{line 25}). Then, it further reduces the search space by identifying buckets containing terms with lower \textit{‘number of tokens‘} (\texttt{line 27}), which is followed by locating buckets having terms with lower \textit{‘term length‘} (\texttt{line 29}), and finally, finding buckets having terms with smaller or equal \textit{‘largest token length`}  (\texttt{line 31}).


{\large \bf Super and Nested Term Retrieval:} \emph{TBI} has $O(1)$ time complexity for super term retrieval (from $Table\_SuperTerms$).  The nested term retrieval complexity is same as that of the hash table (i.e. $O(m^2)$). To retrieve nested terms, candidate nested terms are generated from all possible n-grams of the input term. If a candidate term is not present as a key in $Table\_SuperTerms$, it is discarded. Remaining terms are returned as result.

\section{Data}
We did experiments on 4 different corpora and 1 big dataset (Table \ref{tbl-corpora}). Three of them are from the Penn Treebank-3 dataset \cite{marcus:1999}  -- Brown (1 million words of American English from 1961), WSJ  (1 million words from the 1989 Wall Street Journal) and SWBD (transcribed spontaneous conversations from American English dialects). The NYT corpus (news from New York Times) is obtained the Gigaword corpus \cite{graff:2003}. For these 4 corpora, we used noun phrases (NPs) as terms. They were extracted from using a popular open source NLP tool\footnote{https://spacy.io/}.

The big dataset is Google w2v model \cite{mikolov:2013} containing 3 million terms from a Google News dataset of about 100 billion words.

\section{Empirical Evaluation and Conclusion}

{\large \bf Term Retrieval:} The first set of experiments show that \emph{TBI} is preferable for super and nested term retrieval than Trie based state-of-the-art methods. We compare with \citet{lu:2012} and \citet{aho:1975}. An open source implementation\footnote{https://github.com/robert-bor/aho-corasick} is used for the former, and UMLS Sub-Term Mapping Tools\footnote{https://specialist.nlm.nih.gov/stmt/} for the latter.

%
%

These (Trie) implementations allow only nested term retrieval. So, we enhanced their codes to retrieve super terms. We added an attribute for the nodes in the Trie that keeps track of the maximum depth of the sub-tree for any given node. 
This allows the system(s) to ignore traversing the sub-tree of a node (for a super term match) when number of remaining words (characters, for Aho--Corasick) of the input term, yet to be matched, is bigger than the maximum depth of that sub-tree.

Table \ref{tbl-term-retrieval} shows the average number of node comparisons required per term for the different approaches. Note, a hash table is basically a graph. 
For nested term retrieval, our approach performs almost as well as \citet{lu:2012}. However, for super term retrieval, our approach is vastly superior to either of the other approaches. We skipped these retrieval experiments on the big w2v dataset since the expected results are obvious from the results of other 4 datasets.

{\justify \large \bf Indexing:} A second set of experiments compare indexing performance of TBI with a traditional hash table implementation done by \citet{zhang:2016}.\footnote{All experiments were done on a macOS machine with 2.5 GHz Intel Core i7 processor and 16 GB RAM.} Table \ref{tbl-indexing} shows in each of these dataset, TBI requires substantially, at least 81\%, less time with respect to that of the traditional hash table indexing approach.

{\justify \large \bf Conclusion:} In this paper, we presented a new very fast approach for nested and super term indexing. Unlike commonly used Trie based algorithms, proposed TBI can be used for efficiently retrieving both super and nested terms. We were able to significantly reduce runtime for NLP applications such as terminology and taxonomy extraction using proposed \emph{Terminological Bucket
Indexing} and hope the NLP community will benefit from it, too. 

\begin{table}
\begin{center}
  \begin{tabular}{|c|c|c|}
    \hline
    & \multicolumn{2}{|c|}{Total runtime} \\ \cline{2-3}
    			 & Vanilla hash &  \\ 
    Data & table indexing  & Proposed TBI \\ \hline \hline
    Brown  & 37.18 min & \bf 6.97 min {\it (81\% less)} \\ \hline
    WSJ & 35.18 min & \bf 6.4 min {\it (82\% less)} \\ \hline
    SWBD & 3.57 min & \bf 0.55 min {\it (85\% less)} \\  \hline
    NYT & 1.13 min & \bf 0.18 min {\it (84\% less)} \\  \hline
    Google w2v & 453.02 hours & \bf 65.33 hours {\it (86\% less)} \\  \hline
  \end{tabular}
  \caption{Comparisons of performance between traditional hash table indexing and proposed TBI.}
\label{tbl-indexing}
\end{center}
\end{table}

\begin{algorithm*}[h]
\caption{Proposed indexing algorithm to index a vocabulary inside the proposed term store}\label{index_algo}
\begin{algorithmic}[1]
\Function {$build\_index$}{$list\_of\_terms$}
 \State Let $sorted\_list\_of\_terms$ := terms in $list\_of\_terms$ sorted by term length in ascending order
 \For{Each $new\_term$ in $sorted\_list\_of\_terms$}
  \If {$new\_term$ is not added as a key $Table\_SuperTerms$}
   \State Let $len$ := number of characters (i.e. term length) in $new\_term$
   \State Let $tot\_tok$ := number of tokens in $new\_term$
	\State Let $lt\_len$ := number of characters in the largest token in $new\_term$
	\State Let $tot\_nsp\_char$ := $len$ - $tot\_tok$ + 1
	\State Add a new key $new\_term$ in $Table\_SuperTerms$ with an empty set as value
	\State $insert\_term\_in\_bucket(new\_term, Table\_Bucket, tot\_nsp\_char, tot\_tok, len, lt\_len)$
	\State $update\_nested\_terms(new\_term, Table\_Bucket, tot\_nsp\_char, tot\_tok, len, lt\_len,$ $Table\_SuperTerms)$
  \EndIf
 \EndFor
\EndFunction
\State
\Function {$insert\_term\_in\_bucket$} {$new\_term, Table\_Bucket, tot\_nsp\_char, tot\_tok, len, lt\_len$}
		\State $bucket\_for\_tot\_nsp\_char$ := $Table\_Bucket.get\_or\_initialize\_value$(key=$tot\_nsp\_char$)
		\State $bucket\_for\_tot\_tok$ := $bucket\_for\_tot\_nsp\_char.get\_or\_initialize\_value$(key=$tot\_tok$)
		\State $bucket\_for\_term\_len$ := $bucket\_for\_tot\_nsp\_char.get\_or\_initialize\_value$(key=$len$)
		\State $set\_of\_terms$ = $bucket\_for\_tot\_nsp\_char.get\_or\_initialize\_value$(key=$lt\_len$)
		\State $set\_of\_terms.add$($new\_term$)
\EndFunction		
\State
\Function{$update\_nested\_terms$} {$new\_term, Table\_Bucket, tot\_nsp\_char, tot\_tok, len, lt\_len,$ $Table\_SuperTerms$}
 \For{$k1$ in $Table\_Bucket$.keys() \textbf{where} $k1 \lt tot\_nsp\_char$}
		\State $bucket\_for\_k1$ := $Table\_Bucket.get\_value$(key=$k1$) 
        \For{$k2$ in $bucket\_for\_k1$.keys() \textbf{where} $k2 \lt tot\_tok$}
            	\State $bucket\_for\_k2$ := $bucket\_for\_k1.get\_value$(key=$k2$)
                \For{$k3$ in $bucket\_for\_k2$.keys() \textbf{where} $k3 \lt len$}
                    	  \State $bucket\_for\_k3$ := $bucket\_for\_k2.get\_value$(key=$k3$)
                        \For{$k4$ in $bucket\_for\_k3$.keys() \textbf{where} $k4 \leq lt\_len$}
                                \State $candidate\_nested\_terms$ := $bucket\_for\_k3.get\_value$(key=$k4$)
								  \For{Each $t$ in $candidate\_nested\_terms$ \textbf{where} $new\_term$ is super term of $t$}
								  			\State $Table\_SuperTerms.add\_value$(key\_nested\_term=$t$, value\_super\_term=$new\_term$)
                                \EndFor	
                        \EndFor
                \EndFor	
        \EndFor
 \EndFor
\EndFunction
\end{algorithmic}
\end{algorithm*}

\clearpage
\newpage

\bibliography{nested_term_ref}
\bibliographystyle{acl_natbib}

\end{document}